\documentclass{eas}
\usepackage{graphicx,color}

\newcommand{\et}{et al.\ }
\newcommand{\eg}{e.g., }
\newcommand{\bmu}{\mbox{\boldmath{$\mu$}}}

\newcommand{\bSigma}{\mbox{\boldmath{$\Sigma$}}}
\newcommand{\bTheta}{\mbox{\boldmath{$\Theta$}}}

\newcommand{\bZ}{\mbox{\boldmath{$Z$}}}
\newcommand{\bX}{\mbox{\boldmath{$X$}}}

\begin{document}

\title{The Power of Principled Bayesian Methods in the Study of Stellar Evolution}

\author{Ted von Hippel}
\address{Department of Physical Sciences, Embry-Riddle Aeronautical
University, 600 S. Clyde Morris Blvd, Daytona Beach, FL 32114, USA}

\author{David A. van Dyk}
\address{Statistics Section, Department of Mathematics, Imperial College
London, SW7 2AZ, UK}

\author{David C. Stenning}
\address{Department of Statistics, University of California, Irvine, CA 92617, USA}

\author{Elliot Robinson}
\address{Argiope Technical Solutions, LLC, 816 SW Watson St., Fort White, FL 32038, USA}

\author{Elizabeth Jeffery}
\address{Department of Physics and Astronomy, James Madison University, 901
Carrier Dr, MSC 4502, Harrisonburg, VA 22807, USA}

\author{Nathan Stein}
\address{Statistics Department, The Wharton School, University of
Pennsylvania, 400 Jon M. Huntsman Hall, 3730 Walnut Street, Philadelphia,
PA 19104, USA}

\author{William H. Jefferys}
\address{Department of Astronomy, University of Texas at Austin and
Department of Mathematics and Statistics, University of Vermont, 16
Colchester Ave, Burlington, VT 05401, USA}

\author{Erin O'Malley}
\address{Department of Physics \& Astronomy, Dartmouth College, 6127 Wilder Laboratory,
Hanover, NH 03755, USA}

\begin{abstract}

It takes years of effort employing the best telescopes and instruments to
obtain high-quality stellar photometry, astrometry, and spectroscopy.
Stellar evolution models contain the experience of lifetimes of theoretical
calculations and testing.  Yet most astronomers fit these valuable models
to these precious datasets by eye.  We show that a principled Bayesian
approach to fitting models to stellar data yields substantially more
information over a range of stellar astrophysics.  We highlight advances in
determining the ages of star clusters, mass ratios of binary stars,
limitations in the accuracy of stellar models, post-main-sequence mass
loss, and the ages of individual white dwarfs.  We also outline a number of
unsolved problems that would benefit from principled Bayesian analyses.

\end{abstract}

\maketitle
\runningtitle{Bayesian Studies of Stellar Evolution}

\section{Introduction}

\subsection{Motivation}

Collecting good stellar photometry and spectroscopy is expensive.  For deep
photometry or spectroscopy we use some of the largest ground-based
telescopes, which cost approximately \$1/s to build and operate.  For many
globular cluster studies, we use the Hubble Space Telesope, which is even
more precious.  Likewise, stellar models are expensive.  Careers have been
devoted to understanding stellar structure and evolution and capturing that
knowledge in stellar codes and isochrones.  Yet, typically when astronomers
compare their precious stellar data to these valuable models, they do so by
overplotting theory and data in the same color-magnitude diagrams (CMDs),
then adjusting various parameters until the theory and data appear to fit
best, as judged by eye.  This so called chi-by-eye procedure is neither
strictly reproducible nor does it yield the highest quality results because
the human eye cannot optimize multidimensional parameters or visualize
multidimensional data, each datum of which has its own uncertainties.
Fortunately, there are new statistical solutions available.

In the following sections, we present our Bayesian statistical approach to
this data-fitting problem, and show through a series of examples how this
approach is substantially more capable than the classical, chi-by-eye
approach.  We also outline a number of unsolved problems that would benefit
from Bayesian analysis and point the interested reader toward our
open-source software.

\subsection{Methodological Limitations to Obtaining Scientific Objectives}

When investigators are faced with rich datasets, such as photometry,
spectroscopy, and proper motions or radial velocities, they tend to take a
step-by-step approach to analyzing their data by breaking the problem down
into smaller pieces with different teams working on different apsects of
the problem.  For instance, one team may analyze the spectroscopy to
determine the cluster metallicity, though that in turn depends on the
effective temperature for the stars in question, for which they may return
to the literature for values.  Another team may isolate a sample of stars
with a certain cut-off probability of being cluster members based on radial
velocity and/or proper motions, then publish CMDs along with chi-by-eye
based isochrone fits.  Yet, how are stars to be weighted in these isochrone
fits?  Shouldn't every star have a weight inversely proportional to its
photometric errors and directly proportional to its probability of being a
cluster member, at least above some threshold where the cluster membership
probabilities blend into the larger pool of field stars?  Historically it
is often the case that yet another team performes a more detailed follow-up
study by using the previously-determined metallicity and cluster membership
probabilities to perform a more thorough isochrone fit, perhaps varying
modeling parameters such as convective overshoot or mixing length, though
still fitting models to data by eye.  Yet, such a study may derive a
reddening value for the cluster that is different from the one assumed by
the group deriving cluster spectroscopic metallicity, which ultimately
depended on stellar effective temperature and therefore also on reddening.
Without the spectroscopic material in hand, this study cannot solve for all
of the important parameters simultaneously.

This step-by-step approach can be inconsistnent, relying on different
assumed cluster parameters (most typically, distance and reddening) and
different underlying stellar evolution models.  This approach also assumes
that errors can be propagated by the simple rules of independent normal errors.
Yet stellar evolution is inherently non-linear (\eg stellar lifetimes and
luminosities are not linearly dependent on stellar masses), so there is no
{\it a priori} reason to expect that error distributions are normal.

Besides the general issue of objectivity, there are other limitations in
comparing models to data by eye.  One is that stellar isochrones do not
generally match stars throughout the CMD, and this mismatch can be filter
dependent.  This is clearly visible in the CMDs of NGC 188 presented by
Sarajedini \et (1999, figure 10), where, for instance, a given set of
isochrones matches best around 6 Gyr for a $U-B$ vs.\ $V$ CMD, but appears
older than 7 Gyrs in $B-V$ vs.\ $V$ and $V-I$ vs.\ $V$ CMDs.  Additionally,
in all of the CMDs presented by Sarajedini et al., the isochrones are a few
tenths of a magnitude too blue along the red giant branch (RGB).  How does
one properly judge a match between theory and observations by eye when some
parts of the theory may match better than others?  While we do not have a
complete answer to this question, we believe that the first step in this
direction is to develop an objective technique and demonstrate how it
performs.  A better approach would be to supplement the observational
errors with model errors that captured the uncertainty in stellar evolution
models.  This will be difficult to incorporate until theorists are able to
quantify the uncertainties in their models.

Another question regarding model errors has to do with the proper
adjustment of tunable parameters within a model, \eg mixing length, and how
best to match them to cluster data.  If one does this fit by eye, it is
infeasible to allow all the other parameters to vary; one usually starts by
performing a differential comparison with some best fit. Yet this is
probably sensitive to the starting point, \ie, to whatever the author
thinks is the nominal best fit, and a chi-by-eye fit necessarily focuses on
just a few features in one or maybe two CMDs rathar than simultaneously
fitting all the available photometry.  The problem is compounded if one
wants to generalize this and vary more than one theory ingredient at a
time.  For example, theory ingredients may be coupled or may match the data
in some correlated fashion.

Standard methods also do not allow one to take advantage of ancillary
information such as cluster distances from trigonometric parallaxes or the
moving cluster method, binary mass ratios from radial velocity monitoring,
or white dwarf (WD) masses from fitting the Balmer lines, etc.  Our
Bayesian method allows such ancillary information to be incorporated into
an analysis in a principled manner.

Our technical goal is to avoid as many of these pitfalls as possible and
fully use valuable data.  This means analyzing a star cluster with all the
data we can bring to bear and to consider these data simultaneously via an
objective and rigorous technique.  This technical goal is in service of our
scientific goals, which are 1) to improve our understanding of stellar
evolution by more accurately and precisely comparing models to data, and 2)
deriving better stellar cluster and field star ages.  In Section 2 we
introduce Bayesian statistics, which form the foundation of our statistical
approach.  In Section 3 we introduce our Bayesian software.  In Section 4
we illustrate its application with the dual hope of convincing current
users of chi-by-eye to start using more rigorous approaches and to spark
interest in the development of new imaginative statistical approaches to
the myriad rich datasets that are now available to astronomers.

\section{Bayesian Basics}

\subsection{Introduction}

Thomas Bayes (1702-1761) was a British mathematician and Presbyterian
minister, known for having formulated a special case of Bayes' theorem,
which was published posthumously (Wikipedia, retrieved Jan 12, 2014).
Subsequently, Pierre-Simon Laplace (1749-1827) introduced a general version
of the theorem and used it to approach problems in celestial mechanics,
medical statistics, reliability, and jurisprudence.

\subsection{Bayesian Inference}

Bayes' theorem states that 
\begin{equation}
p(\hbox{model}|\hbox{data}) = \frac{p(\hbox{data}|\hbox{model}) \ p(\hbox{model})}{p(\hbox{data})}
\label{eq:post}
\end{equation}
where $p(\hbox{model}|\hbox{data})$ is the joint (multivariate)
probability density function of a set of model parameters {\it given the
data}, $p(\hbox{data}|\hbox{model})$ is the likelihood function, and
$p(\hbox{model})$ is the prior distribution of the model parameters.  The
prior distribution folds in knowledge from previous analyses of other
datasets, knowledge that one has before considering the current dataset. In
contrast, the posterior distribution, $p(\hbox{model}|\hbox{data})$,
summarizes the combined knowledge for the model parameters stemming both
from the current data and from prior information. The likelihood function
describes the statistical model for the current data set. The value for
$p(\hbox{data})$ is obtained by integrating the numerator over the possible
values of the model parameters so that
\begin{equation}
p(\hbox{model}|\hbox{data}) = \frac{p(\hbox{data}|\hbox{model}) \ p(\hbox{model})}{\int p(\hbox{data}|\hbox{model}) \ p(\hbox{model}) \ d\hbox{model}}. 
\end{equation}
This integral is a normalizing constant that ensures that the posterior
distribution properly integrates to one.  In practice, we use numerical
techniques such as Markov chain Monte Carlo that do not require us to
compute the integral.

\section{BASE-9}

\subsection{Introduction}

We have developed a Bayesian approach to fitting isochrones to stellar
photometry (see \cite{vH06}; \cite{degennaro}; \cite{vanDyk};
\cite{stein}).  Our Bayesian approach is implemented in the software
package called BASE-9\footnote{BASE-9 is available from the first author or
at webfac.db.erau.edu/$\sim$vonhippt/base9.} for {\bf B}ayesian {\bf
A}nalysis of {\bf S}tellar {\bf E}volution with {\bf 9} Parameters.  BASE-9
compares main sequence through asymptotic giant branch stellar evolution
models (\cite{girardi}; \cite{yi}; \cite{dotter}) as well as WD interior
(\cite{wood92}; \cite{montgomery}; \cite{renedo}) and atmosphere models
(\cite{bergeron95}) to photometry in any combination of photometric bands
for which there are data and models.  BASE-9 was designed to analyze star
clusters and accounts for individual errors for each data point; includes
ancillary information such as cluster membership probabilities from proper
motions or radial velocities, cluster distance (\eg from Hipparcos
parallaxes or the moving cluster method), and cluster metallicity from
spectroscopic studies; and can incorporate information such as individual
stellar mass estimates from dynamical studies of binaries or spectroscopic
atmospheric analyses of WDs.  BASE-9 uses a computational technique known
as Markov chain Monte Carlo (MCMC) to derive the Bayesian joint posterior
distribution given in Equation~\ref{eq:post} for six parameter categories
(cluster age, metallicity, helium content, distance, and reddening, and
optionally a parameterized IFMR, see \cite{stein}) and brute-force
numerical integration for three parameter categories (stellar mass,
binarity, and cluster membership).\footnote{Formally, the MCMC sampler
works on the marginalized six-dimensional parameter obtained by numerically
integrating out the three other parameters for each star at each iteration.
Marginalizing in this way substantially improves the computational
performance of the overall MCMC sampler, see \cite{stein} for details.}
The last three of these parameter categories include one parameter per star
whereas the first six parameter categories refer to the entire cluster.  As
a result, for star clusters BASE-9 actually fits hundreds or thousands of
parameters (= 3 N$_{\rm star}$ + 6) simultaneously.

\subsection{Mathematics Behind BASE-9}

The details of the statistical methods we use are described more completely
in \cite{stein}, see also \cite{degennaro} and \cite{vanDyk}.  In
specifying the likelihood function, we use stellar evolution models to
predict observed photometric magnitudes as a function of the several
cluster and stellar parameters, $\bTheta=$ ($m-M$, [Fe/H], log(age),
$A_{\rm V}$, stellar mass, etc.).  These predictions incorporate
main-sequence/red-giant models, white dwarf models, a parameterized
initial-final mass relationship, and the possibility of unresolved binary
stars.  Specifically, letting $\bmu_i(\bTheta)$ be the predicted photometric
magnitudes for star $i$, the likelihood function is
$$\small
L(\bTheta,\bZ | \bX, \bSigma) =
 \prod_{i=1}^{n}
\Bigg[ 
{Z_i\over \sqrt{(2\pi)^{n}|\bSigma_i|}}
 \exp\left(-{1\over2}\Big(\bX_i - \bmu_i(\bTheta)\Big)^\top\bSigma_i^{-1}\Big(\bX_i -\bmu_i(\bTheta)\Big) \right) 
$$
 \begin{equation} \small
 \qquad\qquad + \ \  (1- Z_i) p_{\rm field}(\bX_i)
\Bigg] ,
\label{eq:like}
\end{equation}
where $n$ is the number of stars in the dataset, $\bX =(\bX_1,\ldots,
\bX_n)$ are the observed photometric magnitudes, $\bSigma =
(\bSigma_1,\ldots, \bSigma_n)$ are the known variance-covariance matrices
of the observed magnitudes, $\bZ =(Z_1,\ldots, Z_n)$ are unknown zero-one
variables that indicate whether each star is a cluster star $(Z_i=1)$ or a
field star $(Z_i=0)$, and $p_{\rm field}$ is the distribution of
photometric magnitudes for field stars and is assumed uniform in the range
of the data.

Treating the denominator of Equation~\ref{eq:post} as a normalizing
constant, the posterior distribution in our Bayesian analysis can be
written, namely
\begin{equation}
p(\bTheta,\bZ |  \bX, \bSigma) \propto L(\bTheta | \bX, \bSigma) p(\bTheta,\bZ),
\end{equation}
where $p(\bTheta,\bZ)$ is the prior distribution for the stellar parameters
and of the cluster-member indicator variables.  Computationally, BASE-9
works on the marginal posterior distribution of  Equation~3.2 obtained by
summing over $\bZ$ and numerically integrating over the star-specific
masses.

The prior distribution quantifies the knowledge and uncertainty we have
about the stellar parameters before considering the current data set.  The
posterior distribution combines this information with that in the current
data, updating our knowledge about the stellar parameters in light of the
new information.

\section{Applications}

\subsection{Cluster Properties from Main Sequence Turn-off Stars and White Dwarfs}

The age for the halo of the Milky Way is derived from main sequence
turn-off (MSTO) fitting of globular clusters, whereas the age of the disk
of the Milky Way is derived from white dwarf models fit to the white dwarf
luminosity function.  Because both of these age determinations are based on
different and largely unrelated physics, we need to confirm whether they
are on the same absolute age scale.  The best way to accomplish this is to
study single-age star clusters and derive both their MSTO ages and their WD
cooling ages.  Yet, to take full advantage of this approach, one must take
great care to derive MSTO and WD ages consistently.

\setlength{\abovecaptionskip}{-5pt} 
\setlength{\belowcaptionskip}{-10pt} 

% Figure 1
\begin{figure}[!ht]
\centering
\includegraphics[width=\textwidth]{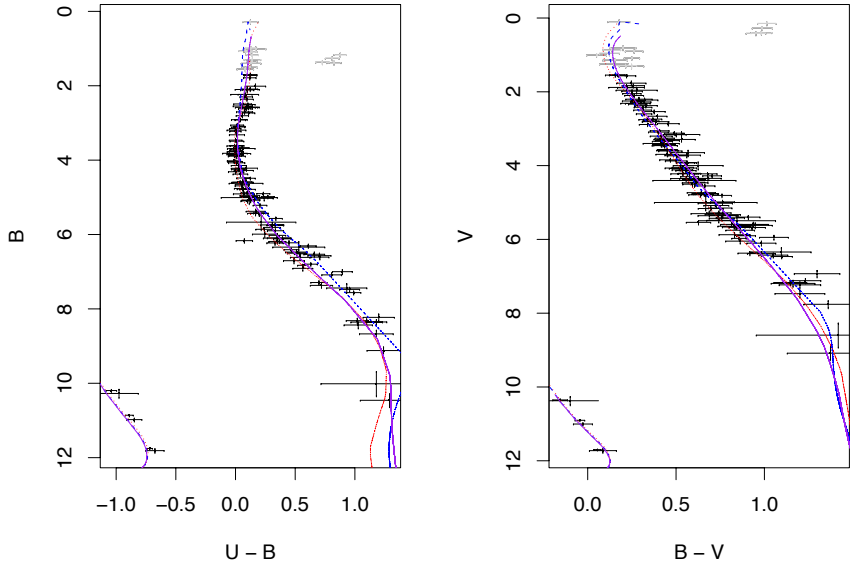}
\caption{Isochrones for three different main-sequence model sets at the
nominal age, distance, and metallicity of the Hyades.  Solid purple lines
represent the \cite{girardi} models, dotted red lines represent the
\cite{yi} models, and dashed blue lines represent the \cite{dotter} models.
Gray points show additional Hyades photometry not used in these fits.
(Figure from \cite{degennaro}.)} 
\end{figure}

% Figure 2
\begin{figure}[!h]
\centering
\includegraphics[width=\textwidth, angle=0, scale=0.85]{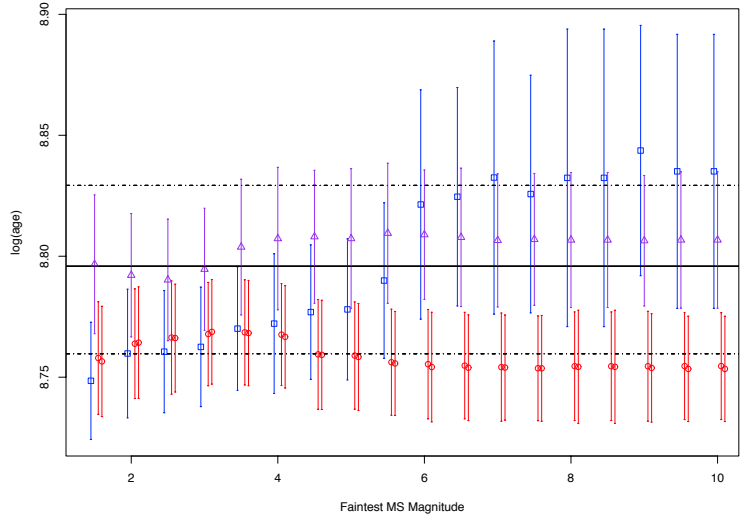}
\caption{Derived log10(age) for the Hyades as a function of the
faintest main sequence star included in the analysis.  (Figure from
\cite{degennaro}.)}
\end{figure}

% Figure 3
\begin{figure}[!h]
\centering
\includegraphics[width=\textwidth, angle=0, scale=0.85]{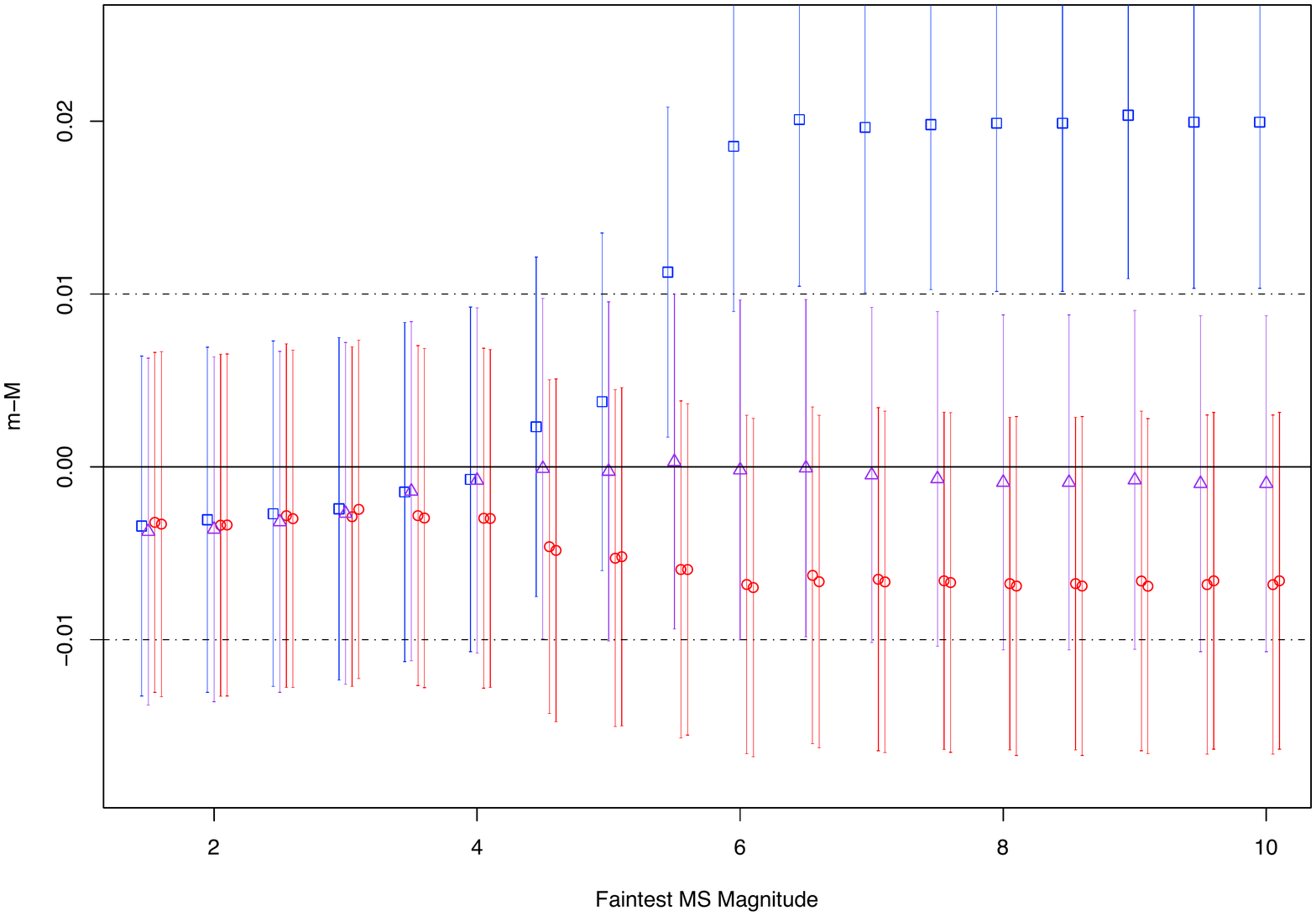}
\caption{Derived Hyades distance modulus relative to \cite{perryman} as
a function of the faintest main sequence star included in the analysis.
(Figure from \cite{degennaro}.)} 
\end{figure}

Figure 1, from \cite{degennaro}, displays Hyades $UB$ and $BV$ CMDs along
with overplotted isochrones from \cite{girardi}, \cite{yi}, and
\cite{dotter}.  These isochrones represent fits that simultaneously
incorporate the cluster main sequence and WDs.  These fits do not include
the MSTO or red giant branch (plotted in gray), as the goal of this
particular study was to derive the cluster WD age.  Yet, it is clear from
this figure that each of these stellar evolution models fails to fit the
lower main sequence, with different offsets for each model.  None of these
three sets of models claims to fit low mass stars, yet how does this
mis-fitting propagate into creating a best fit isochrone?  \cite{degennaro}
circumvent this problem by studying the best fit as a function of depth
along the main sequence, with more stars being included further down the
main sequence in each of a succession of fits.

In Figure 2 we present the derived log10(age) for the Hyades as a function
of the faintest main sequence star included in the analysis from
\cite{degennaro}.  The blue squares represent \cite{dotter} models, the red
circles represent two different runs of the \cite{yi} models (to test
sensitivity to starting values), and the purple triangles represent the
\cite{girardi} models.  The error bars on the symbols span the 68\%
probability interval in the log10(age) posterior distribution.  The
horizontal lines are the mean and $\pm 1 \sigma$ deviations of the most
reliable estimate for the age of the Hyades as determined by MSTO fitting
by \cite{perryman}.  In Figure 3 we present the offset in the derived
distance modulus for the Hyades as a function of the faintest main sequence
star included in the analysis of \cite{degennaro} relative to that found by
\cite{perryman}.  The symbols have the same meaning as in Figure 2.  The
distance values comes from \cite{perryman} and represents the prior
information used in the analysis.  We can see that in both figures most
fits agree with previous values for the Hyades until $V \approx 8$, where
we see larger departures.  In the case of the Hyades, the most relevant
constraint from prior work is the Hipparcos distance published by
\cite{perryman}, and we can see the offset is well within the very small
errors.

% Figure 4 (7.1 of Jeffery (2009) PhD Thesis)
\begin{figure}[!h]
\includegraphics[width=\textwidth, angle=270, scale=0.68]{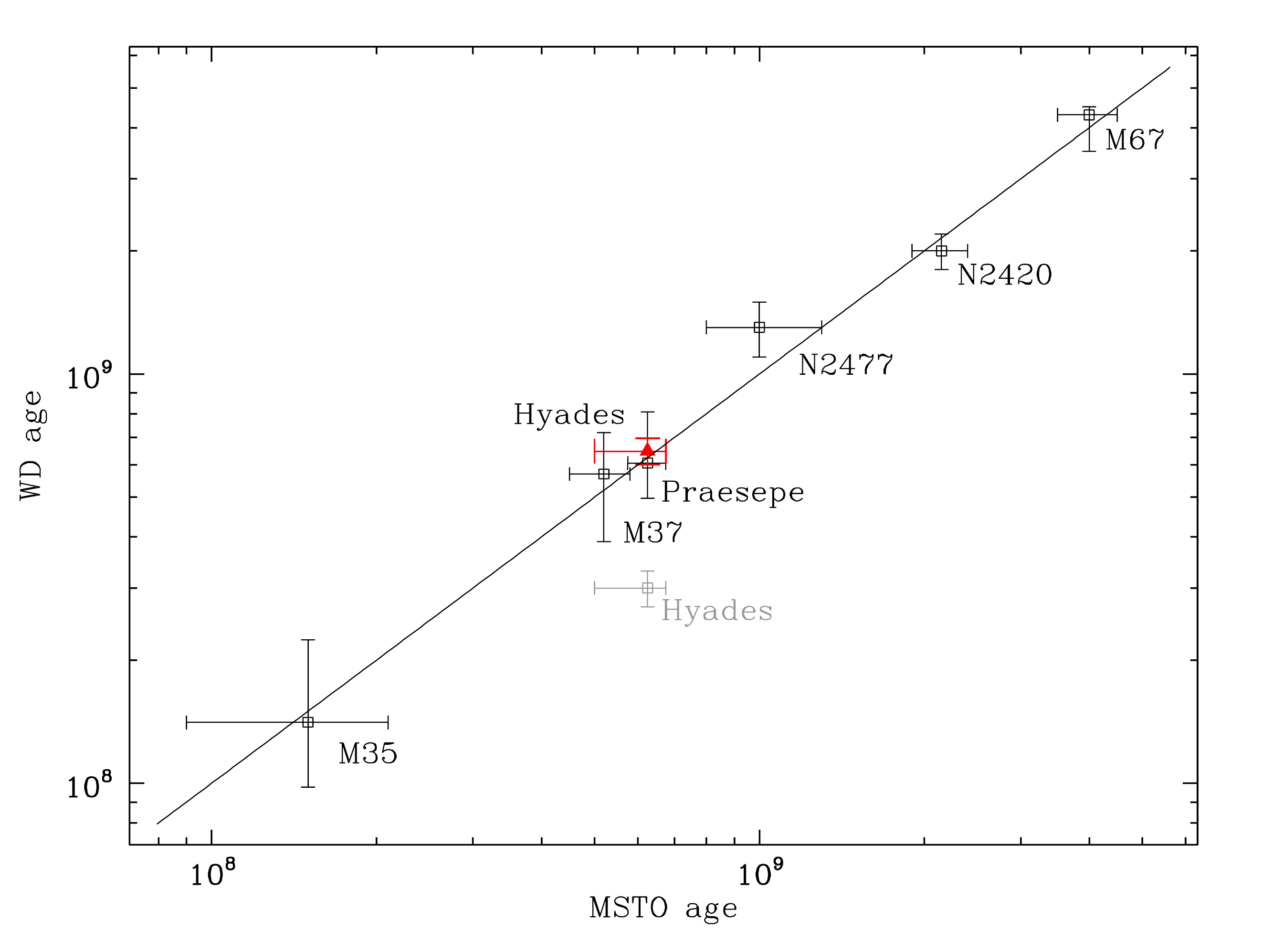}
\caption{WD versus MSTO age for seven clusters. (Figure from
\cite{jeffery09}.)} 
\end{figure}

\cite{jeffery09} continues this study and finds that the MSTO and WD ages
are consistent to the limit of their study at 4 Gyr.  Figure 4 shows the WD
versus MSTO age for seven clusters from \cite{jeffery09}.  The age derived
from the WDs in the Hyades by \cite{degennaro} using our Bayesian approach
brings the Hyades age into agreement with the MSTO age for the first time
(red solid triangle). The solid line shows the one-to-one correspondence
between the WD and MSTO ages, and the gray point shows the most reliable WD
age of the Hyades prior to the work by \cite{degennaro}.

% Figure 5 (5.7 of Jeffery (2009) PhD Thesis)
\begin{figure}[!h]
\includegraphics[width=\textwidth, angle=270, scale=0.7]{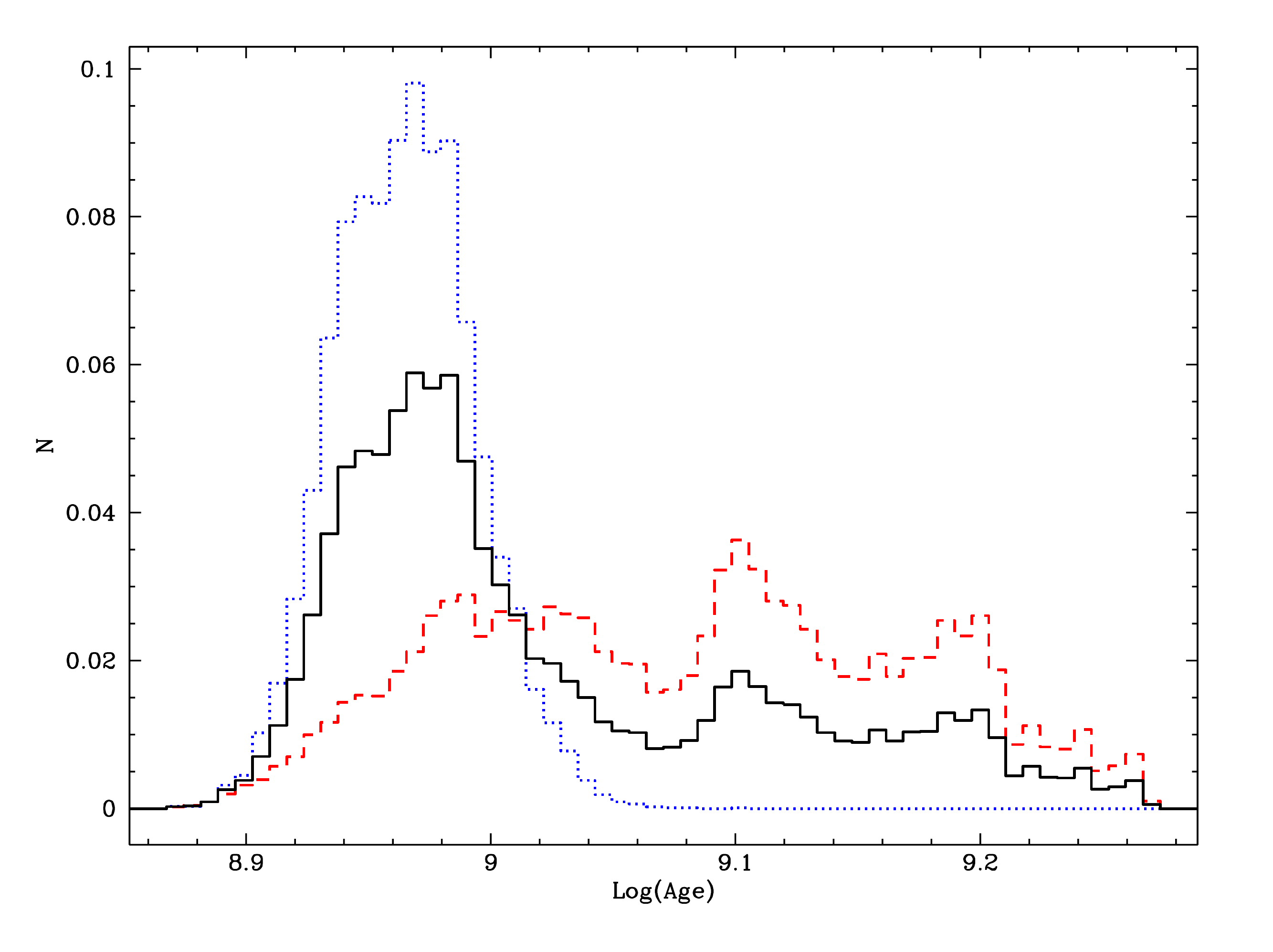}
\caption{The age distribution of NGC 2360 with and without the inclusion of
one particular WD.  (Figure from \cite{jeffery09}.)} 
\end{figure}

Perhaps more demonstrative of the power of the Bayesian approach is Figure
5 from \cite{jeffery11}, which shows the shapes of age posterior
distributions and their sensitivities to individual stars.  The solid black
line is the complete age posterior distribution when all WD candidates in
NGC 2360 are included.  The dotted blue line is the age distribution when
WD5 is included as a cluster member.  The red dashed line is the
distribution when WD5 is excluded as a field star. This indicates the
importance of this star in determining the location of the WD cooling
sequence and hence measuring the cluster age.

\subsection{Cluster Binary Stars}

Figure 6 presents the simulated effects of unresolved binaries on the CMD.
The black line is an isochrone from the \cite{yi} models, which indicates
the position of single stars.  The complex hooks emanating from the
isochrone show the effect of adding secondary stars to primary stars at
specific points along the isochrone. The binary pairs closest to the main
sequence isochrone have a secondary star with a mass of 0.4 $M_{\odot}$ and
each successive point further away from the main sequence is the result of
a higher mass secondary star until the final point 0.75 mag above the
isochrone, where the binary consists of two equal mass stars.  Figure 7
presents an enlargement of the area in Figure 6 around the three proximal
binary sequences, including also an example photometric data point (in red)
with errors.  The photometry can match a number of different primary and
secondary stars.

% Figure 6
\begin{figure}[!h]
\includegraphics[width=\textwidth, angle=0, scale=0.85]{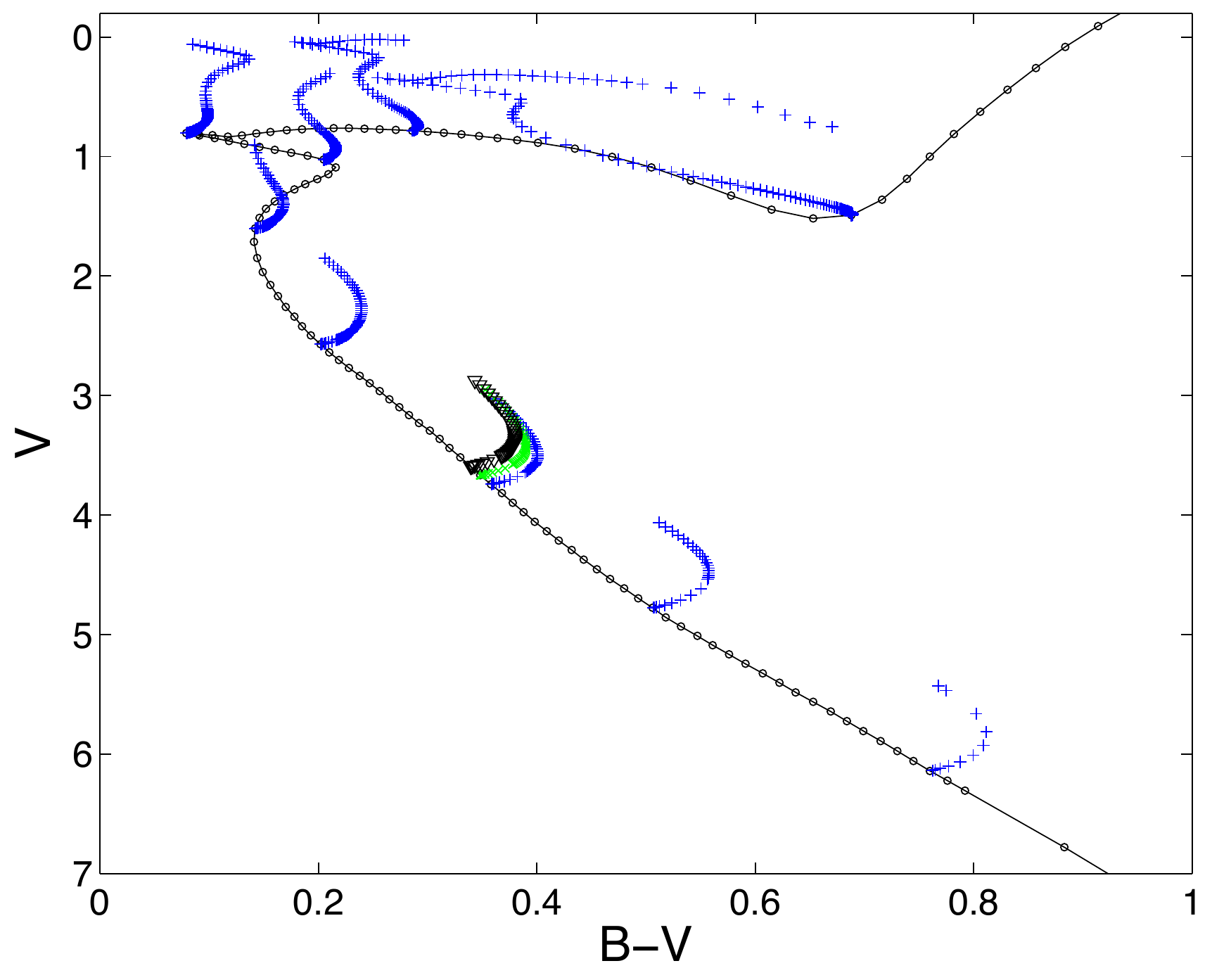}
\caption{An example of the effects of unresolved binaries on the CMD.}
\end{figure}

% Figure 7
\begin{figure}[!h]
\includegraphics[width=\textwidth, angle=0, scale=0.87]{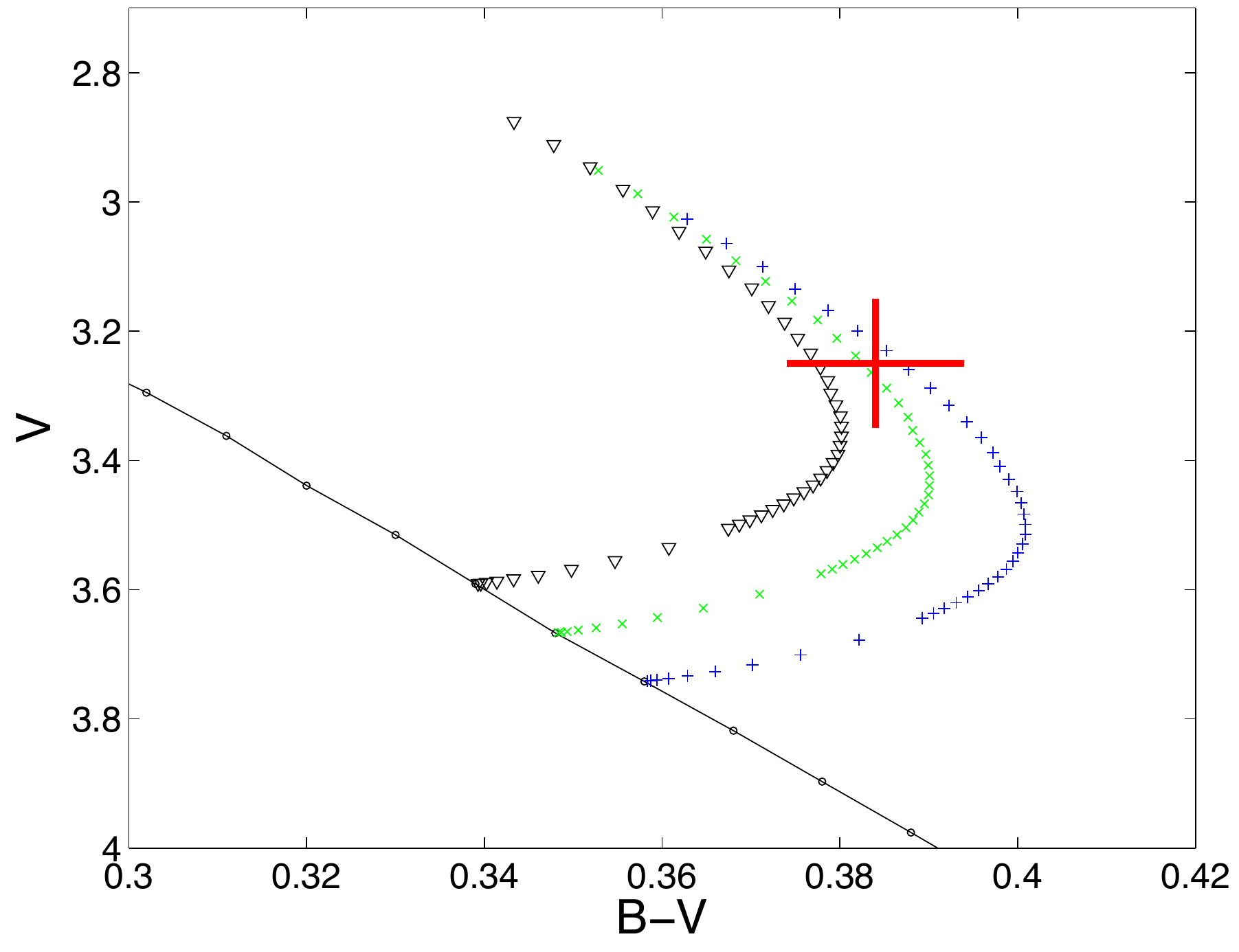}
\caption{Enlargement of the area around the three proximal binary
sequences, including also an example photometric data point (in red) with
errors.}
\end{figure}

Figure 8, an updated version from \cite{vanDyk}, resents the joint
posterior distribution for the primary and secondary masses of the star
vB022 in the Hyades.  The scatter plot shows the Monte Carlo sample from
the posterior distribution using the three stellar evolution models
discussed above and presented in Figures 1-3.  The star has a posterior
probability of cluster membership equal to 99.955\% and the plot gives the
conditional posterior distribution of the two masses given that the binary
system is a member of the cluster.  This is compared with a kinematic
estimate of the two masses based on fitting radial velocities to this
elicpsing double-lined spectroscopic binary and is indicated by the point
with $\pm$ 1$\sigma$ error bars.  The primary mass is marginally
inconsistent ($\sim$2$\sigma$) and slightly lower ($\sim$ 0.02 to 0.03
$M_{\odot}$) than the more reliable external estimate.  (All masses are in
units of $M_{\odot}$.) These types of comparisons can provide tests of
stellar evolution models.

% Figure 8
\begin{figure}[!h]
\includegraphics[width=\textwidth, angle=0, scale=0.8]{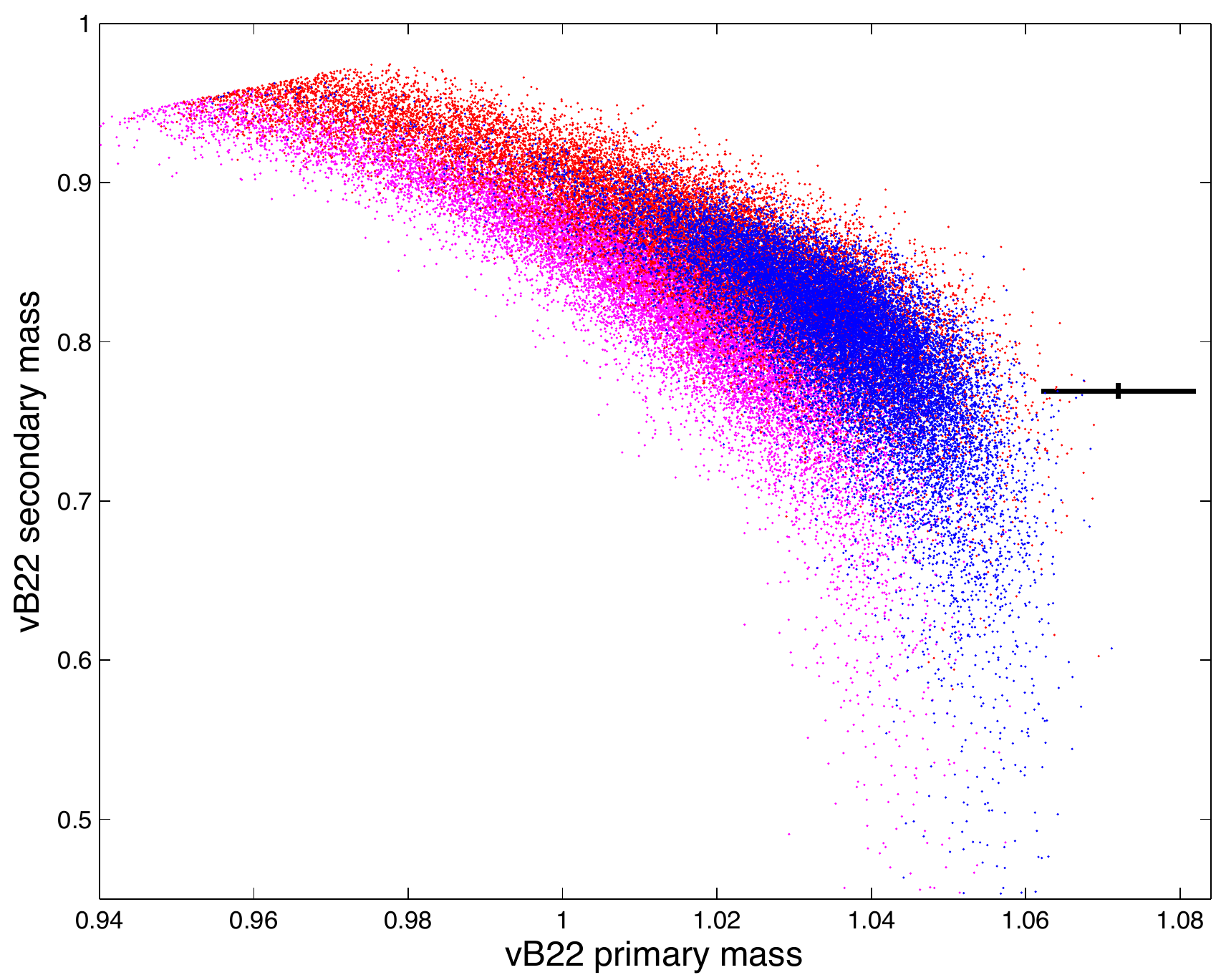}
\caption{The joint posterior distribution for the primary and secondary
masses of the Hyades binary vB022.  (Figure from \cite{vanDyk}.)}
\end{figure}

\subsection{Sensitivity Analysis: Results Depend on Filters}

\cite{hills} showed that the cluster parameters one derives can depend on
the filters one choses.  This both demonstrates limitations in the stellar
evolution models and is a cautionary tale to anyone who might compare
cluster ages (or other parameters) fit from different filter photometry.
Figure 9 shows CMDs for NGC 188 after the cluster was cleaned of most
binaries.  The overplotted isochrone is a fit based on \cite{yi} models
applied to $UBVRI$ photometry.  The point types correspond to the assigned
cluster membership probability for each star (filled circles have
membership probabilities greater than 0.9, open circles indicate between
0.7 and 0.9 probability, open squares show probabilities less than 0.7).
These probabilities result from the arithmetic mean of membership
probabilities from \cite{stetson} (based on proper motions) and
\cite{geller} (based on radial velocities).

% Figure 9
\begin{figure}[!h]
\centering
\includegraphics[width=\textwidth, scale=0.5]{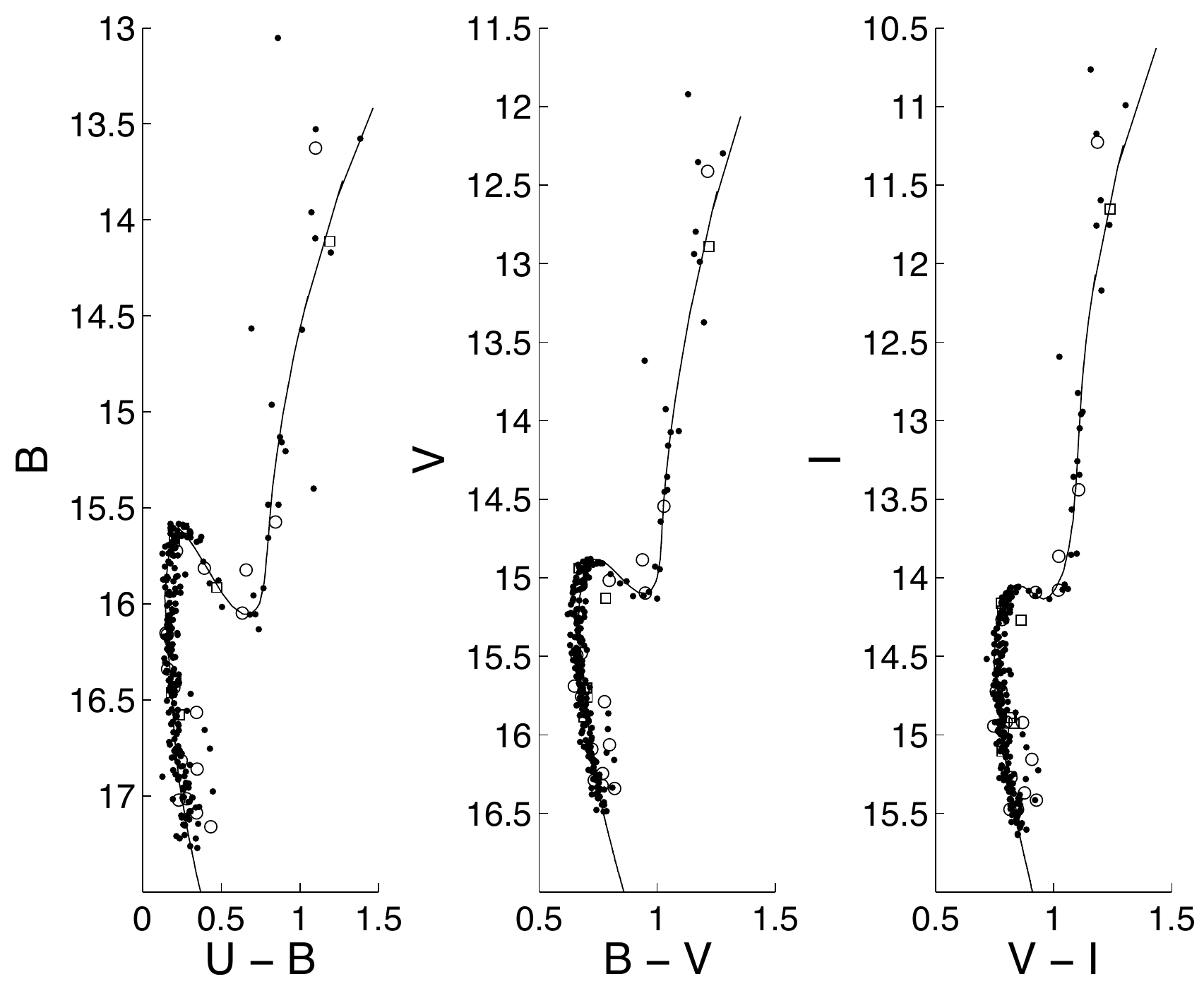}
\caption{CMDs for NGC 188 in a range of optical bands.  (Figure from
\cite{hills}.)}
\end{figure}

% Figure 10
\begin{figure}[!h]
\centering
\includegraphics[width=\textwidth, scale=0.5]{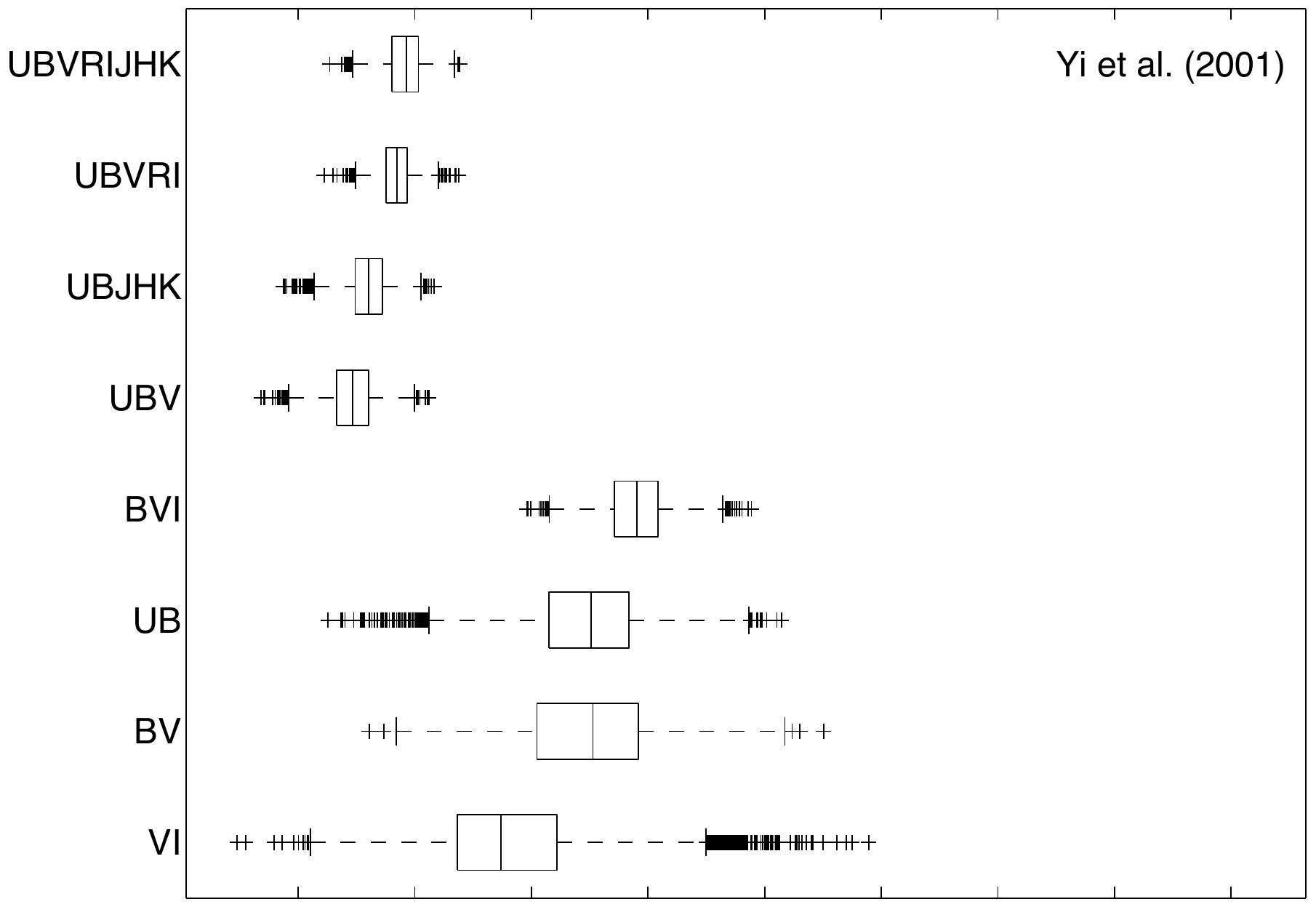}
\caption{Box-and-whisker plots for age across eight photometric band
combinations based on \cite{yi} models.  (Figure from \cite{hills}.)}
\end{figure}

Figure 10 shows box-and-whisker plots from \cite{hills} of age fits across
eight photometric band combinations based on \cite{yi} models.  The
box-and-whisker plots are a compact way of showing the degree to which a
distribution is non-Gaussian.  The central line indicates the median of the
distribution and the upper and lower box edges indicate the 25th and 75th
percentiles.  The whiskers extend out to the most extreme non-outliers, and
outliers are plotted individually. A data point is considered an outlier if
it is smaller than $q_1 - \frac{3}{2}(q_3 - q_1)$ or greater than $q_3 +
\frac{3}{2}(q_3 - q_1)$, where $q_1$ and $q_3$ are the 25th and 75th
percentiles, respectively.  Some filter combinations yield meaningfully
different results, and many of these are inconsistent at the equivalent
of 3$\sigma$.

\subsection{Initial Final Mass Relation}

Mass lost by stars ascending the red giant branch is poorly understood (see
\cite{catelan}).  At the same time, red giant mass loss is key to
interpreting globular cluster and old open cluster CMDs because it drives
the morphology of the horizontal branch (HB).  In more distant systems,
such as M31, HB morphology may be our only clue to the ages of stellar
systems.  Yet this clue is fraught with uncertainty because we lack a
properly determined mass loss formula for stellar models.  A proper
understanding of mass loss becomes even more important beyond the Local
Group where we interpret stellar populations only through their integrated
light.

Work to date has focused either on theoretical (\eg \cite{blocker}) or
semi-empirical (\eg \cite{reimers}; \cite{judge}) parameterizations of this
process or on constraints from the remnant WD masses (\eg \cite{weidemann};
\cite{williams04}; \cite{kalirai05}).  Yet these approaches have been
necessarily piecemeal, with either a focused examination of the theoretical
constraints (\eg \cite{vassiliadis}) or an iterative and not
self-consistent approach to deriving the mass lost from the mains sequence
to the WDs (see \cite{salaris} for a thorough critical evaluation).  In
addition, these mass loss studies are based on only a few key star clusters
(\eg \cite{origlia}) because quality data have been so limited.

An independent way to measure mass loss during the late stages of stellar
evolution is to determine, via cluster ages, the mapping of initial zero
age main sequence (ZAMS) masses to final WD masses, which is the so-called
Initial-Final Mass Relation (IFMR). With the exception of Omega Cen and a
few other globular clusters, star clusters are essentially or at least
approximately single-age systems.  As such, the age derived from the WDs
should be equal to the age derived from the MSTO stars. The former
incorporates the age of the WD precursor, the IFMR itself, and then WD
cooling rates. Such work has been done in the past (e.g., \cite{weidemann};
\cite{kalirai05}; Williams, Bolte, \& Koester (2004); Williams, Bolte, \&
Koester (2009)), though always in the context of a step-wise solution,
often in an internally inconsistent manner (\cite{salaris}).  Because
stellar evolution is a highly non-linear process, the step-wise approach
does not adequately propagate errors, even when done consistently, and it
certainly does not recover the entire posterior distribution on any
parameterization of the IFMR. Within the Bayesian context, one can solve
this problem with complete internal consistency for each stellar evolution
model and recover the IFMR parameter distribution.  In Figure 11 we show
this type of analysis performed by \cite{stein}.  In this example,
\cite{stein} set the IFMR to be a simple linear model with two free
parameters (its slope and intercept) and then used BASE-9 to fit and assess
uncertainty in these two IFMR parameters while simultaneously fitting all
the other cluster and stellar parameters.  While the IFMR is unlikely to be
linear, any given cluster constrains just a small mass range of this
relation, and a linear fit is appropriate within this small range.
\cite{stein} also performed fits using broken-linear and quadratic IFMRs.

% Figure 11 - Fig 4 from Stein et al.
\begin{figure}[htp]
\centering
\includegraphics[width=\textwidth]{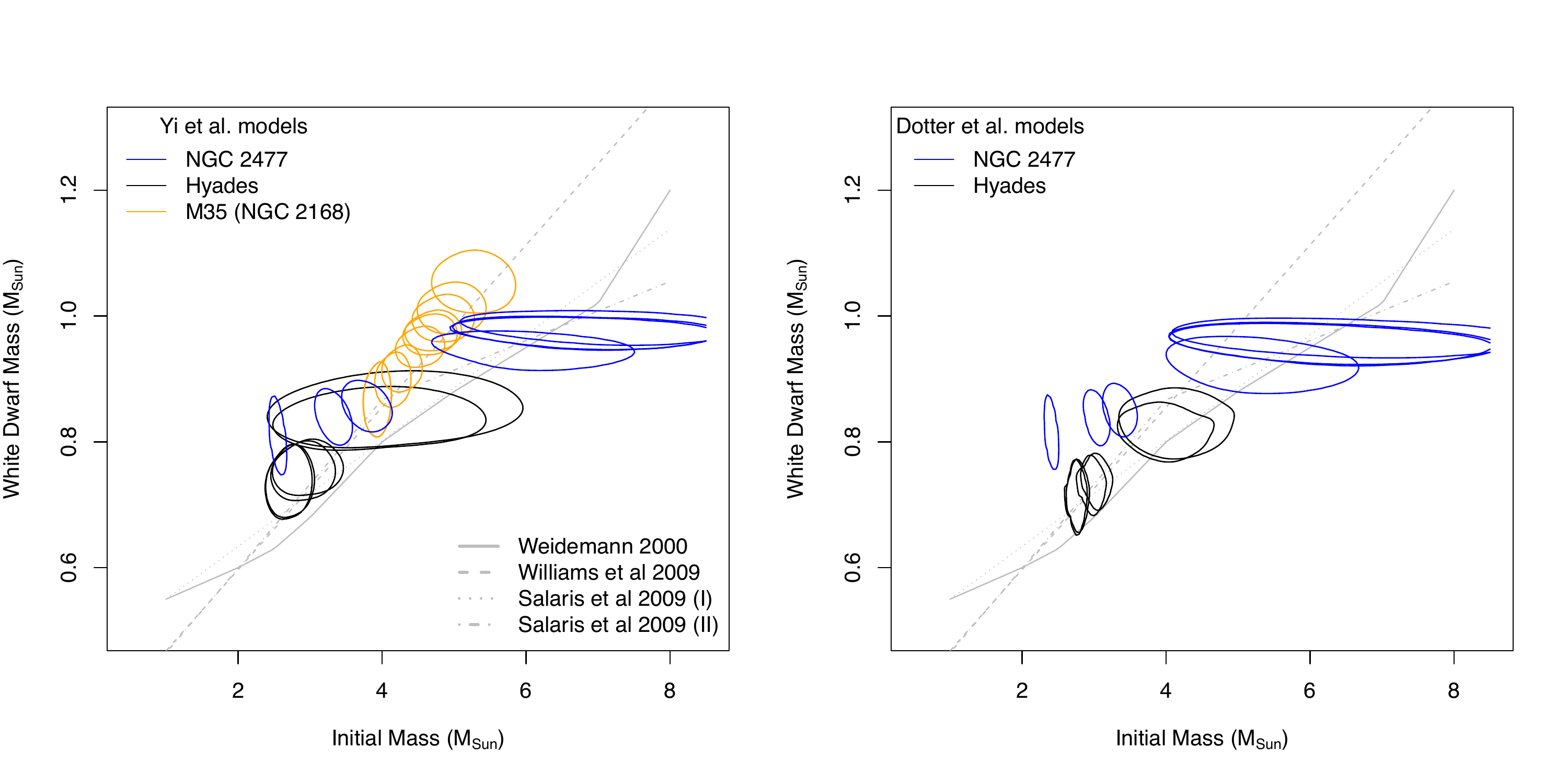}
\caption{The initial and WD masses for WDs from the open clusters NGC 2477,
the Hyades, and M35 analyzed using the \cite{yi} models.  Contours are 95\%
highest posterior density regions of each WD's joint initial and final mass
posterior distribution.  In this particular analysis, the IFMR model
enforces linearity within each cluster, but not between clusters.  (Figure
from \cite{stein}.)}
\end{figure}

\subsection{Ages for Individual Field White Dwarfs}

While one cannot typically derive ages for individual main sequence or red
giant branch stars, one can derive ages for individual WDs because WDs have
a mass-radius relation that constrains their luminosity.  In many cases,
this constraint is sufficient to yield useful ages.  (In a few years, we
expect GAIA astrometric results to open up new possibilities for
determining the ages of individual stars.)

\cite{BLR} improved the technique for deriving individual WD ages by
comparing WD masses and $T_{\rm eff}$ values with WD cooling models.  This
technique relies on measuring $T_{\rm eff}$ from photometry or spectroscopy
and log(\emph{g}) from spectroscopy or WD surface area from trigonometric
parallax.  Because WDs have a mass-radius relation (\cite{hamada}), either
log(\emph{g}) or surface area yield mass, and the mass and $T_{\rm eff}$,
when compared to a WD cooling model, yield the WD cooling age.  The WD mass
is relied upon again to infer its precursor mass through the
imprecisely-known IFMR.  The precursor mass is then converted to a pre-WD
lifetime via stellar evolution models.  Finally, the precursor lifetime is
added to the WD cooling age to determine the total age of the WD.

The above-mentioned technique has its advantages and disadvantages.  The
foremost advantage is that it yields reasonably precise ages for individual
WDs.  Secondarily, for cool WDs with masses $\geq$ 0.7 M$_{\odot}$, the
progenitor lifetime is short relative to the WD cooling age, and therefore
uncertainties in the IFMR are unimportant (see, for instance, \cite{vH06},
fig 16).  On the negative side, this age technique involves many steps,
some of which are often performed inconsistently.  An additional negative
is that one has to correctly propagate errors through many steps.  Some
errors may start out symmetrically distributed (\eg $T_{\rm eff}$), but the
assumptions behind the standard propagation of errors are not met
(see \cite{omalley}), casting doubt on the estimates and errors that this
approach produces.

% Figure 12, which is Fig 3 of O'Malley, von Hippel, & van Dyk (2013)
\begin{figure}[!h]
\centering
\includegraphics[width=\textwidth]{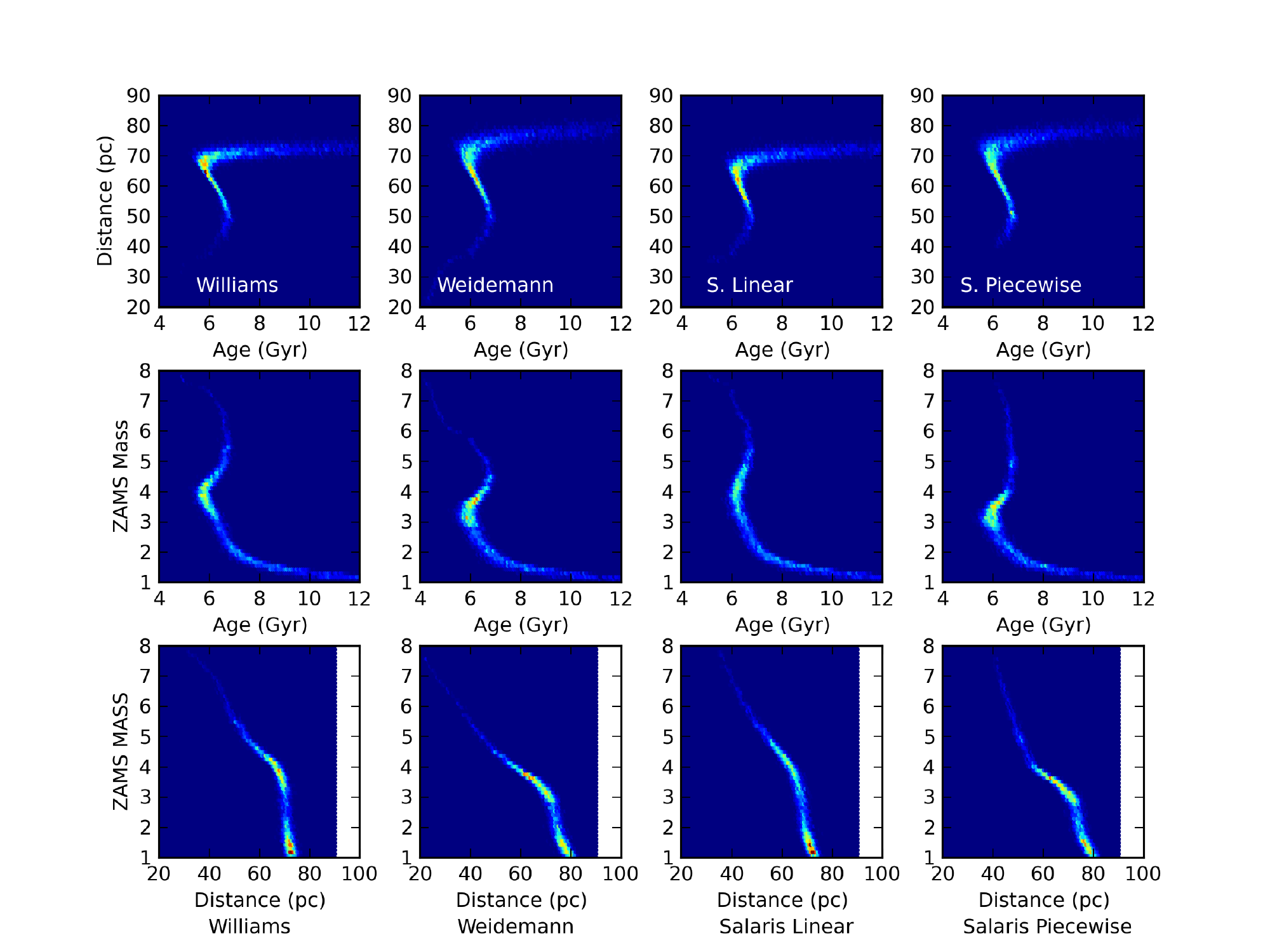}
\caption{Posterior probability projections in the age, distance, ZAMS mass
planes for the WD J0003$-$0111 for each of four IFMRs and the
\cite{montgomery} WD cooling models.  These IFMRs, from left to right, are
the \cite{williams09}, \cite{weidemann}, and \cite{salaris} Linear and
Piecewise relations.  (Figure from \cite{omalley}.)}
\end{figure}

Figure 12 displays the posterior distributions for a single representative
WD (J0003$-$0111), fit with the \cite{montgomery} WD models and each of
four IFMRs studied by \cite{omalley}.  There are detailed similarities in
all four cases, and in fact the contours for all IFMRs peak near 6 Gyrs and
65 pc, yet the distributions are subtly different from one IFMR to another.
\cite{omalley} showed such comparisons for 28 WDs and were able to make a
detailed study of the effect of different IFMRs on implied WD ages.

\subsection{Future Applications}

In addition to the subjects studied above, there are a number of other
problems in stellar astrophysics that we believe would benefit from a
principled Bayesian analysis.  A short list of examples follow.

\subsubsection{Multiple populations in globular clusters}

While it has been known for decades that some globular clusters contain
stars with a range of abundances, it has only recently become clear that
globular clusters can host multiple stellar populations.  Exquisite Hubble
Space Telescope photometry now clearly shows multiple main sequences,
subgiant branches, or red giant branches in a number of clusters.  This
newly recognized complexity challenges current chi-by-eye approachs far
more than past work on globular clusters, both because more parameters are
involved in any fit and because newly abundant spectroscopic data tags
individual stars with additional information indicating which population
they likely belong to and some details about the properties of that
population.

\subsubsection{Carbon-to-oxygen ratio in WDs}

Stellar evolution models predict the carbon-to-oxygen ratio as a function
of stellar mass but these calculations are affected by uncertainties in the
nuclear reaction rates, overshooting, and mass loss.  Thus, it is important
to check these results empirically both as a test of stellar evolution
models and because the carbon-to-oxygen ratio affects the WD cooling rate,
and thereby the implied age of any observed WD. The density-temperature
phase diagram for carbon+oxygen mixtures has not been empirically verified
at these densities and temperatures and thus crystallization and the
possibility of phase separation of C and O remain as possible uncertainties
in deriving ages for the oldest WDs, particularly those in globular
clusters.  One can use Bayesian techniques to compare families of WD models
to the two or three globular clusters and eight open clusters with
sufficient WD populations and perform simultaneous age fits for both the
WDs and MSTO stars, with the carbon-to-oxygen ratio as a free parameter.

\subsubsection{Width of the main sequence}

What is the intrinsic width of the main sequence once one accounts for
photometric errors, binaries, and other known effects?  These other effects
include rapid stellar rotation (\eg \cite{mermilliod}; \cite{james}) and
stellar activity (\eg \cite{pace}), both of which are prevalent in young
clusters.  In some clusters, differential reddening broadens the main
sequence (\eg NGC 2477, \cite{vonHippel95}).  For clusters where these
effects can be minimized or properly modeled, a careful Bayesian analysis
has the sensitivity to test for internal metallicity dispersions as small
as $\sigma$([Fe/H]) = 0.02.

There are two reasons to perform this test.  The first is to measure or
place new limits on the degree of gas heterogeneity in the parent clouds
that formed today’s open clusters and globular clusters.  This in turn
would measure the ratio of the turbulent mixing time scale to the length of
time the cloud existed before forming stars.  Such a metallicity
heterogeneity would cause a broadening of the main sequence largely
independent of stellar mass.

The second reason to perform this measurement is to test whether there are
detectable signatures from the ingestion of planetary material during giant
planet migration.  This latter effect should be noticeable for stars with
shallow convection zones such as F stars. \cite{pins} find that stars
hotter than $T_{\rm eff}$ = 6700 would have metallicities elevated by 0.02
dex after ingesting only one earth mass of iron.  \cite{trilling} argue
that it is likely that more planets are accreted onto their host star than
remain in orbit, and thus if 10\% of all solar type stars have planets we
might expect an even greater fraction to have ingested planetary material.
Unlike parent cloud heterogeneity, this effect would be mass dependent and
thus we could distinguish the two possible sources.  Measuring stellar
photospheric abundance variations within a cluster could provide insight
into the ubiquity and ultimate fate of planetary migration.

\subsubsection{Tests of physical ingredients of stellar evolution}

Convective core overshoot is a crucial ingredient in stellar models of
stars with masses greater than about 1 solar mass, hence ages less than a
few Gyr.  In general, overshoot mixing applies to all stellar models with
convective boundaries but the most dramatic evidence for this is in the
convective core where overshoot mixing draws H-rich material into the core,
thereby extending the main sequence lifetime of the star.

Hydrodynamic simulations of convection have been performed for a number of
environments (\eg \cite{freytag}; \cite{meakin}) with the end result that
it is possible to incorporate the improved understanding of overshoot
mixing gained from these simulations into stellar evolution codes.  This
treatment of convective overshoot has been investigated in detail in models
of AGB stars (\cite{herwig97}; \cite{herwig00}) but has not been
systematically applied to the CMD morphology of young and intermediate age
open clusters.  Current cluster photometry, combined with a principled
Bayesian approach, can test whether or not the hydrodynamically-motivated
overshoot mixing prescription is able to reproduce the CMD morphologies of
various open clusters with a range of age and metallicity.

\subsubsection{Coupling age information from multiple field WDs}

Following on the results of \cite{omalley}, it would be particularly useful
if we could extract Galactic population age information from disk, thick
disk, and halo populations of WDs.  In many cases, we will have
probabilistic population asignments based on proper motions and distances,
yet there are other constraints such as a common IFMR, at least for a given
metallicity, and different metallicity distributions for the precursor
populations.  These constraints, combined with the observational data,
need to be combined in a simultaneous fit to recover the age distribution
of our Galaxy's major components.

\section{Conclusions}

We have used a number of examples to show that a principled Bayesian
approach to fitting stellar data with models yields substantially more
information with substantially higher precision than the classic chi-by-eye
approach.  We have demonstrated that our Bayesian technique yields more
precise and accurate star clusters properties, mass ratios of binary stars,
and ages of individual white dwarfs.  We have further demonstrated that our
technique illuminates limitations in the accuracy of stellar isochrones and
quantifies post-main-sequence mass loss.  There are a wide number of
additional topics in stellar evolution that will benefit from principled
Bayesian analysis.

\acknowledgements
We would like to thank the organizers of the Ecole Every Schatzmann for
making it possible for TvH, DS, and EO to attend and contribute to this
school.  This material is based upon work supported by the National
Aeronautics and Space Administration under Grant NNX12AI54G issued through
the Office of Space Science. In addition, DvD was supported in part by a
British Royal Society Wolfson Research Merit Award, by a European
Commission Marie-Curie Career Integration Grant, and by the STFC (UK).

%% bibliography

\end{document}